\documentclass[conference]{IEEEtran}

\usepackage[utf8]{inputenc}
\usepackage[cmex10]{amsmath}
\usepackage{graphicx}
\usepackage{amssymb}
\usepackage{amsthm}
\usepackage{subfigure}
\usepackage{cite}
\usepackage{color}

\usepackage{xcolor}

\theoremstyle{remark}

\usepackage[ruled,norelsize]{algorithm2e}
\makeatletter
\newcommand{\removelatexerror}{\let\@latex@error\@gobble}
\makeatother
\usepackage{algorithmicx}

\begin{document}

\title{RACE: A Rate Adaptive Channel Estimation Approach for Millimeter Wave MIMO Systems}

\IEEEoverridecommandlockouts 
\author{
    \IEEEauthorblockN{Matthew Kokshoorn, He Chen, Yonghui Li, and Branka Vucetic}
    \IEEEauthorblockA{        
        School of Electrical and Information Engineering, The University 	of Sydney, Australia \\
        Email: \{matthew.kokshoorn, he.chen, yonghui.li,  branka.vucetic\}@sydney.edu.au }
    \thanks{This research was supported by ARC grants DP150104019 and FT120100487. The research was also supported by funding from the Faculty of Engineering and Information Technologies, The University of Sydney, under the Faculty Research Cluster Program and the Faculty Early Career Researcher Scheme.}
}

\maketitle

\begin{abstract}
In this paper, we consider the channel estimation problem in Millimeter wave (mmWave) wireless systems with large antenna arrays. By exploiting the inherent sparse nature of the mmWave channel, we develop a novel rate-adaptive channel estimation (RACE) algorithm, which can adaptively adjust the number of required channel measurements based on an expected probability of estimation error (PEE). To this end, we design a maximum likelihood (ML) estimator to optimally extract the path information and the associated probability of error from the increasing number of channel measurements. Based on the ML estimator, the algorithm is able to measure the channel using a variable number of beam patterns until the receiver believes that the estimated direction is correct. This is in contrast to the existing mmWave channel estimation algorithms, in which the number of measurements is typically fixed. Simulation results show that the proposed algorithm can significantly reduce the number of channel estimation measurements while still retaining a high level of accuracy, compared to existing multi-stage channel estimation algorithms. 
\end{abstract}

\section{Introduction}

Recently, the Millimeter wave (mmWave) spectrum has been regarded as a promising alternative to the congested microwave frequencies used in today's cellular networks \cite{pi2011introduction,rappaport2013millimeter,hong2014study}. In the past, this spectrum has been largely underutilized owing to its severe signal propagation loss compared to that over conventional frequencies \cite{rheath,zhang2010channel,torkildson2010channel}. Recently it has been shown that the small mmWave wavelength  enables the integration of a large number of antennas at both the transmitter and the receiver to form a narrow high gain beam to compensate the propagation loss. However, effective beamforming relies heavily on an accurate estimation of channel state information.

The channel estimation for mmWave systems is non-trivial for a number of reasons. The first challenge stems from the sampling requirement of large bandwidth signals expected in mmWave systems. With a large number of antennas, it becomes impractical to equip every antenna with an expensive RF chain and an ADC with high sampling rate \cite{pi2011introduction}. To reduce both hardware cost and power consumption, several works have proposed analog beamforming techniques \cite{5284444,beamcodebook,chen2011multi}. The fundamental idea of analog beamforming is to control the phase of the received signals on each antenna via a network of analog phase shifters which are connected to a single ADC. Similar hardware reductions can be made at the transmitter in terms of digital to analog conversion. The second challenge is owing to the signal propagation loss inherent to the  mmWave channel. Because of this loss, before beamforming, omni-directional transmission of any pilot signal results in very low receive signal-to-noise ratios (SNRs), leading to very inaccurate estimation. As such, existing mmWave channel estimation algorithms have been typically limited to beamforming based approaches. 

\begin{figure}[!t]
\centering
\includegraphics[width=3.3in,trim={2.1cm 3.5cm 1.0cm 4.5cm},clip]{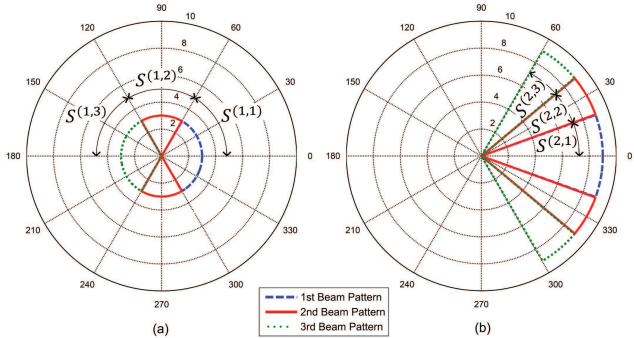}
\caption{Illustration of the beam patterns adopted in the first (a) and second (b) stages of the channel estimation algorithm of \cite{rheath} when $K = 3$. The three sub-ranges in the first stage are, $[0, \pi/3)$, $[\pi/3, 2\pi/3)$ and $[2\pi/3, \pi)$, respectively. By assuming that the possible AOAs/AODs are reduced to the sub-range $[0, \pi/3)$ in the first stage, this sub-range is further divided into $[0, \pi/9)$, $[\pi/9, 2\pi/9)$ and $[2\pi/9, \pi/3)$, respectively, in the second stage.}
\label{beam_patterns}
\end{figure}

On the other side, recent measurements \cite{rappaportMeasure,Akdeniz} have shown that the mmWave channel exhibits sparse propagation characteristics in the angular domain. That is, there are only a few dominant propagation paths in mmWave channels \cite{sayeed2002deconstructing}. Therefore, the key objective of mmWave channel estimation is actually to identify these paths so that the transceiver can align the transmit and receive beams along the identified directions. Leveraging the previously described constraints and channel model, compressed sensing based channel estimation algorithms have been proposed to significantly reduce the estimation measurement overhead by exploiting the channel sparsity in mmWave systems \cite{Kokshoorn,rheath,Compressed_Channel_Sensing}. Many of these existing approaches use `divide and conquer" type algorithms to progressively refine the possible angular range of an angle of departure (AOD) and angle of arrival (AOA) in a multiple stage process. An example of such approach is depicted in Fig. \ref{beam_patterns}. These algorithms can reduce the channel estimation time for each AOD/AOA pair to a logarithmic complexity with respect to the number of antennas.

However, the aforementioned multi-stage channel estimation approaches have an inherent error propagation issue. That is, if the estimated AOD/AOA pair is incorrect in any given stage, all subsequent estimation stages will definitely be incorrect. Further adding to this problem, the loss of directivity gain for beam patterns used at early stages will lead to a low received SNR. As a result, there is a higher probability of error at early stages, where the success of the estimation is at its most critical. In \cite{rheath}, the authors proposed to allocate power among the stages inversely proportional to the beamforming gains, resulting in an equal probability of error in each stage. While this scheme may work well in theory, in practice such approach may require hardware that tolerates a very large peak-to-average power ratio, as the power requirement in the first stage may be an order of magnitude greater than the final stage.

As an alternative approach, more directional beam patterns can be used in the earlier stages, essentially allocating more time to the earlier stages in order to decrease the probability of estimation error. However, increasing the  number of measurements in any stage without knowing whether more are needed, may lead to an unnecessary waste of both time and energy, especially when the channel is in good condition. Ideally, we seek a channel estimation algorithm that can adaptively adjust the number of measurements according to the channel condition. To the best  knowledge of the authors, no such adaptive channel estimation approach for mmWave communication systems exists in open literature.
%

Motivated by this open problem, we aim to develop an adaptive mmWave channel estimation algorithm, which can adapt the number of channel measurements to the channel conditions without knowing the instantaneous channel state information, according to the probability of channel estimation error. In order to implement such an approach, we need an estimator that is capable of satisfying two important properties. First, we need to be able to optimally estimate the AOD/AOA from an increasing number of measurements. The second property is that we are able to estimate the probability of estimation error after each measurement so that we can determine whether more measurements are needed. Due to its optimal and statistical nature, we develop a maximum likelihood estimator to satisfy these requirements.





The key contributions of this paper are summarized as follows. We propose a novel rate-adaptive channel estimation (RACE) algorithm, where the average number of channel measurements is adapted to the channel conditions. To this end, we measure the channel using a variable number of beam patterns until the receiver believes that the estimated AOD/AOA is correct. We design a maximum likelihood (ML) estimator to optimally extract the path information and the corresponding probability of error from a variable number of measurements. When we apply our approach to existing multi-stage estimation techniques, the number of measurements required in each stage can also be dynamically allocated among stages. This results in the allocation of more time to early stages, as opposed to more power. Numerical results demonstrate that the proposed RACE scheme requires substantially less channel measurements compared to the existing approaches with a rate-switching scheme, which alters the number of beam patterns used in early stages based on the expected probability of error in a predetermined manner.

\textit{Notation} : We use capital bold-face letter $\boldsymbol{A}$ to denote a matrix, $\boldsymbol{a}$ to denote a vector, ${a}$ to denote a scalar, and $\mathcal{A}$ denotes a set. $||\boldsymbol{A}||_2$ is the 2-norm of $\boldsymbol{A}$, $\text{det}(\boldsymbol{A})$ is the determinant of $\boldsymbol{A}$. $\boldsymbol{A}^T$, $\boldsymbol{A}^H$ and $\boldsymbol{A}^*$ are the transpose, conjugate transpose and conjugate of $\boldsymbol{A}$, respectively. For a square matrix $\boldsymbol{A}$, $\boldsymbol{A}^{-1}$ represents its inverse. $\boldsymbol{I}_N$ is the $N\times N$ identity matrix and $\lceil \cdot \rceil$ denotes the ceiling function. $\mathcal{C}\mathcal{N}(\boldsymbol{m},\boldsymbol{R})$ is a complex Gaussian random vector with mean $\boldsymbol{m}$ and covariance matrix $\boldsymbol{R}$, and $\text{E}[\boldsymbol{a}]$ and $\text{Cov}[\boldsymbol{a}]$ denote the expected value and covariance of ${\boldsymbol{a}}$, respectively.

\section{System Model}

Consider a mmWave multiple-input multiple-output (MIMO) system composed of $N_t$ transmit antennas and $N_r$ receive antennas. We consider that both the transmitter and receiver are equipped with a limited number of RF chains. Following \cite{rheath}, we further assume that these RF chains, at one end, can only be combined to form a single beam pattern, indicating that only one pilot signal can be transmitted and received at one time. As this serves as the most hardware constrained case, it is then straightforward to consider multiple measurements at one time. To estimate the channel matrix, the transmitter sends a pilot signal $x$, with unit energy ($||x||_2=1$), to the receiver. Denote by $\boldsymbol{f}$ and $\boldsymbol{w}$  $(||\boldsymbol{f}||_2 = ||\boldsymbol{w}||_2 = 1)$, respectively, the $N_t \times 1$ beamforming vector at the transmitter and $N_r \times 1$ beamforming vector at the receiver. The corresponding channel output can be represented as
\begin{equation} \label{y}
y = \sqrt{P} \boldsymbol{w}^H \boldsymbol{H} \boldsymbol{f} x + \boldsymbol{w}^{H} \boldsymbol{q},
\end{equation}
where $\boldsymbol{H}$ denotes the $N_r\times N_t$ MIMO channel matrix, $P$ is the transmit power and $\boldsymbol{q}$ is an $N_r \times 1$ complex additive white Gaussian noise (AWGN) vector following the distribution $\mathcal{C}\mathcal{N}(0, N_0 \boldsymbol{I}_{N_r})$. 

In this paper, we follow \cite{Sayeed_max} and adopt a two-dimensional (2D) sparse geometric-based channel model. Specifically, we consider a single path\footnote{For the purpose of exploration, a single path model is adopted. Note that the proposed RACE scheme could be extended to multi-path scenarios by following a similar way as in [\citenum{rheath}, Algorithm 2].} channel between the transceiver with steering AOD, $\phi^t$, and AOA, $\phi^r$. Then the corresponding channel matrix can be expressed in terms of the physical propagation path parameters as
%
\begin{equation} \label{H_sum}
\boldsymbol{H} = \alpha  \sqrt{N_tN_r} \boldsymbol{a}_{r}(\phi^r) \boldsymbol{a}_{t}^{H}(\phi^t)
\end{equation} 
where $\alpha$ is the fading coefficient of the propagation path, and $\boldsymbol{a}_t(\phi^t)$ and $\boldsymbol{a}_r(\phi^r)$ respectively denote the transmit and receive spatial signatures of the propagation path. To simplify the analysis, we assume that the transmitter and receiver have the same number of antennas (i.e., $N_t = N_r = N$). However, it is worth pointing out that the developed scheme can be easily extended to a general asymmetric system. If uniform linear antenna arrays (ULA) are employed at both the transmitter and receiver, we can define $\boldsymbol{a}_t(\phi^t)= \boldsymbol{u}(\phi^t)$ and $\boldsymbol{a}_r(\phi^r)= \boldsymbol{u}(\phi^r)$, respectively, where 
%
\begin{equation} \label{u_n}
\boldsymbol{u}(\epsilon) \triangleq \frac{1}{\sqrt{N}} [1,e^{j 2 \pi \epsilon},\cdots,e^{j2 \pi (N-1)\epsilon}]^T.
\end{equation}
Here, the steering angle, $\phi^t$, is related to the physical angle $\theta^t \in [0,\pi)$ by $\phi^t=\frac{d\text{ sin}(\theta^t)}{\lambda}$ with $\lambda$ denoting the signal wavelength\footnote{Note that the use of ULA results in no distinguishable difference between AODs $\theta^t$ and $-\theta^t$ or between AOAs $\theta^r$ and $-\theta^r$. Hence, only AODs and AOAs in the range $[0,\pi)$ need to be considered.}. A similar expression can be written for $\phi^r$ at the receiver. With half-wavelength spacing, the distance between antenna elements becomes $d=\lambda/2$.

From (\ref{H_sum}), we can see that the overall channel state information of each path includes only three parameters, i.e., the AOD $\phi^r$, the AOA $\phi^r$, and the fading coefficient $\alpha$. We assume that the fading coefficient follows a complex Gaussian distribution with zero mean and variance $P_R$, and that both $\phi^t$ and $\phi^r$ can only take some discrete values from the set $\mathcal{U}_N=\{0,\frac{\pi}{N},\cdots,\frac{\pi(N-1)}{N} \}$. We aim to find an efficient way to estimate these parameters. The key challenge here is how to design a sequence of $\boldsymbol{f}'s$ and $\boldsymbol{w}'s$ in such a way that the channel parameters can be quickly and accurately estimated.
	
\section{Proposed Method}	

\subsection{Multi-stage Channel Estimation}
We begin by adopting a similar multi-stage approach as \cite{rheath}  (see Fig. \ref{beam_patterns}) where, in each stage $s$, we consider $K$ channel estimation beam patterns that span $K$ angular sub-ranges at both the transmitter and receiver. Denote by $\boldsymbol{f}_k^{(s)}$ and $\boldsymbol{w}_k^{(s)}$, respectively, the transmit and receive beamforming vectors adopted for transmission over the $k$th angular sub-range in the $s$th stage such that $||\boldsymbol{f}_k^{(s)}||_2=||\boldsymbol{w}_k^{(s)}||_2 = 1,\;\forall\;k,s$. We then use $K^2$ time slots to span all possible sub-range combinations, after which we can obtain a sequence of $K^2$ measurements represented as
\begin{align} \label{y_h_v}
\boldsymbol{y}^{(s,K^2)} =& \sqrt{P} x \boldsymbol{h}_v^{(s,K^2)} +\boldsymbol{n}^{(s,K^2)} ,
\end{align} 
where $\boldsymbol{h}_v^{(s,K^2)} $ describes the channel input-output relationship between the $K^2$ transmit and receive beamforming vector combinations defined as
\begin{align} \label{h_v}
    \boldsymbol{h}_v ^{(s,K^2)} =
    \left[\begin{array}{ccc}
   	(\boldsymbol{w}_1^{(s)})^H \boldsymbol{H} \boldsymbol{f}_1^{(s)}  \\
   	(\boldsymbol{w}_2^{(s)})^H \boldsymbol{H} \boldsymbol{f}_1^{(s)}  \\  
   	\vdots			\\
   	(\boldsymbol{w}_1^{(s)})^H \boldsymbol{H} \boldsymbol{f}_2^{(s)}  \\
   	(\boldsymbol{w}_2^{(s)})^H \boldsymbol{H} \boldsymbol{f}_2^{(s)}  \\    	
   	\vdots												\\ 	
   	(\boldsymbol{w}_K^{(s)})^H \boldsymbol{H} \boldsymbol{f}_K^{(s)} 
    \end{array} \right]
\end{align}   
and
\begin{align} \label{n_q}
    \boldsymbol{n}^{(s,K^2)}  =
    \left[\begin{array}{ccc}
   	(\boldsymbol{w}_1^{(s)})^H \boldsymbol{q}_1   \\
   	\vdots									\\ 	
   	(\boldsymbol{w}_K^{(s)})^H \boldsymbol{q}_{K^2}  
    \end{array}\right]
\end{align} 
is a $K^2\times 1$ vector of the corresponding noise terms. Note that  since $||\boldsymbol{w}_k^{(s)}||_2 = 1$, $\forall$ $k$, the vector $\boldsymbol{n}^{(s,K^2)}$ follows the same distribution as that of ${\boldsymbol{q}_m}$, i.e., $\boldsymbol{n}^{(s,K^2)} \sim \mathcal{C}\mathcal{N}(0, N_0 \boldsymbol{I}_{K^2})$, where $\boldsymbol{I}_{K^2}$ is the $K^2\times K^2$ identity matrix.

We design each of the $K$ beamforming vectors based upon $K$ angular sub-ranges. Denote by $\mathcal{S}^{(1,k)}$ the $k$th sub-range of the first stage defined by $\mathcal{S}^{(1,k)}=\{\epsilon | \epsilon \in \mathcal{U}_N| (k-1) \pi /K \leq \epsilon < k \pi /K \}$. More generally, we denote by $\mathcal{S}^{(s,k)}$ the $k$th sub-range in the $s$th stage. We then design each beamforming vector to match the transmit/receive spatial signature in each of these sub-ranges. To this end, the beamforming vector should satisfy the following equation
\begin{equation} \label{matrix_eq_1}
\boldsymbol{U}^H \boldsymbol{f}_k^{(s)} = \boldsymbol{U}^H \boldsymbol{w}_k^{(s)} 
     \triangleq \boldsymbol{z}^{(s,k)}
\end{equation}
where $\boldsymbol{U} = \big[ \boldsymbol{u}(0),\boldsymbol{u} \Big( \frac{\pi}{N} \Big),\cdots,\boldsymbol{u} \Big( \frac{\pi(N-1)}{N} \Big) \big]$ is a matrix whose columns describe the transmit/receive spatial signatures at each angle and
\begin{equation} 
\label{z_out} 
\boldsymbol{z}^{(s,k)}_i = 
\begin{cases}
    C_{(s)},& \text{if } \frac{i \pi}{N} \in \mathcal{S}^{(s,k)} \\
    0,              & \text{otherwise} 
\end{cases}
\end{equation}
where $C_{(s)}$ is a scalar constant that ensures $||\boldsymbol{f}_k^{(s)}||_2=||\boldsymbol{w}_k^{(s)}||_2=1$. From (\ref{matrix_eq_1}), $\boldsymbol{f}_k^{(s)}$ and $\boldsymbol{w}_k^{(s)}$ can be obtained by performing the left pseudo inverse of $\boldsymbol{U}$ giving
\begin{align}
\label{f_design}
\boldsymbol{f}_k^{(s)} = \boldsymbol{w}_k^{(s)} = (\boldsymbol{U}\boldsymbol{U}^H)^{-1}\boldsymbol{U} \boldsymbol{z}^{(s,k)}.
\end{align}

In \cite{rheath}, after the $K^2$ measurements are carried out in each stage, the magnitudes of the $K^2$ measurements are compared. The transmitter and receiver sub-ranges corresponding to the strongest measurement, respectively denoted by $\hat{k}_t$ and $\hat{k}_r$, are selected for estimation in the next stage. The receiver feeds back the value of $\hat{k}_t$ to the transmitter and further pilot signals are limited to the reduced sub-ranges $\mathcal{S}^{(s,\hat{k}_t)}$ and $\mathcal{S}^{(s,\hat{k}_r)}$, by dividing each into a further $K$ sub-ranges for the estimation in the next stage. This process continues until the minimum angle resolution $\frac{\pi}{N}$ is reached. This requires $S=\lceil \text{log}_K(N) \rceil$ stages. Using an example of $K=3$, Fig. \ref{beam_patterns} shows the beam patterns adopted in a two-stage example scenario.

\subsection{Maximum Likelihood Estimation of AOD/AOA Information}
In order to realise the rate adaptive channel estimation, the receiver needs to optimally identify the most likely path sub-ranges from an increasing number of received channel measurements and calculate the probability of  estimation success. To do so, we will develop a maximum likelihood estimator to extract the AOD/AOA information from the $K^2$ measurements. 

From (\ref{y_h_v}) we can express the distribution of $\boldsymbol{y}^{(s,K^2)}$ as
\begin{align} \label{y_dist_1}
    \boldsymbol{y}^{(s,K^2)}\sim\mathcal{C}\mathcal{N}(& \sqrt{P}x \; \text{E}[\boldsymbol{h}_v^{(s,K^2)}]  + \text{E}[\boldsymbol{n}^{(s,K^2)}] , \nonumber \\ & P||x||_2^2 \; \text{Cov}[\boldsymbol{h}_v^{(s,K^2)}]  + \text{Cov}[\boldsymbol{n}^{(s,K^2)}] ).
\end{align}
In order to simplify (\ref{y_dist_1}), recall that $\boldsymbol{n}^{(s,K^2)}\sim \mathcal{C}\mathcal{N}(0,N_0 \boldsymbol{I}_{K^2})$, the pilot signal has unit power $||x||_2^2=1$ and that the path coefficient $\alpha$ has zero mean, resulting in $\text{E}[ \boldsymbol{h}_v^{(s,K^2)}] = 0$. We can then re-write the distribution of $\boldsymbol{y}^{(s,K^2)}$ as
\begin{align} \label{y_dist}
    \boldsymbol{y}^{(s,K^2)}&\sim\mathcal{C}\mathcal{N}(0,P \; \text{Cov}[\boldsymbol{h}_v^{(s,K^2)}]  + N_0 \boldsymbol{I}_{K^2} ) \nonumber \\
    &= \mathcal{C}\mathcal{N}(0, \boldsymbol{\Sigma} )  
\end{align}
where $\boldsymbol{\Sigma}=P \text{Cov}[\boldsymbol{h}_v^{(s,K^2)}]  + N_0 \boldsymbol{I}_{K^2}$ is the covariance matrix of $\boldsymbol{y}^{(s,K^2)}$. As it is difficult to find an expression for $\text{Cov}[\boldsymbol{h}_v^{(s,K^2)}]$, due to the unknown channel, we instead aim to find an expression for the co-variance, conditional on an AOD and AOA being present between the $k_t$th and $k_r$th sub-range, respectively. From (\ref{h_v}), we can re-write a measurement between the $k_t$th transmitter and the $k_r$th receiver sub-range by substituting the channel from (\ref{H_sum}) as
\begin{align} \label{measurement}
(\boldsymbol{w}_{k_r}^{(s)})^H \boldsymbol{H} \boldsymbol{f}_{k_t}^{(s)}=\alpha N (\boldsymbol{u}^H(\phi^r) \boldsymbol{w}_{k_r}^{(s)})^{H}   \boldsymbol{u}^{H}(\phi^t) \boldsymbol{f}_{k_t}^{(s)}.
\end{align}
By observing the output for each beamforming vector in (\ref{matrix_eq_1}-\ref{z_out}), we can then see that the channel measurement in (\ref{measurement}) has two possible outcomes corresponding to whether a path is present between the two sub-ranges or not. More formally, this can be expressed as
\begin{equation} \label{measurment}
(\boldsymbol{w}_{k_r}^{(s)})^H \boldsymbol{H} \boldsymbol{f}_{k_t}^{(s)} =
\begin{cases}
    \alpha N C^2_{(s)},& \text{if } \phi^t \in \mathcal{S}^{(s,k_t)} \text{ and } \phi^r \in \mathcal{S}^{(s,k_r)} \\
    0,               & \text{otherwise}. \\
\end{cases}
\end{equation}
Then, by assuming transmit and receive sub-range estimates $\hat{k}_t$ and $\hat{k}_r$ satisfy $ \phi^t \in \mathcal{S}^{(s,\hat{k}_t)} \text{ and } \phi^r \in \mathcal{S}^{(s,\hat{k}_r)}$, we can define the conditional co-variance as 
\begin{align} \label{sigma}
\boldsymbol{\Sigma}_{(\hat{k}_t,\hat{k}_r)} &= P \; \text{Cov}[\boldsymbol{h}_v^{(s,K^2)} |\hat{k}_t,\hat{k}_r  ]  + N_0 \boldsymbol{I}_{K^2}
\end{align}
where the conditional $\boldsymbol{h}_v^{(s,K^2)}$ can be found by substituting (\ref{measurment}) into (\ref{h_v}), and the abbreviated condition `$|\hat{k}_t,\hat{k}_r$' represents `$|\phi^t \in \mathcal{S}^{(s,\hat{k}_t)}, \phi^r \in \mathcal{S}^{(s,\hat{k}_r)}$'. The conditional co-variance term can then be calculated by $\text{Cov}[\boldsymbol{h}_v^{(s,K^2)} |\hat{k}_t,\hat{k}_r  ] = \text{E}[\boldsymbol{h}_v^{(s,K^2)} (\boldsymbol{h}_v^{(s,K^2)})^H |\hat{k}_t,\hat{k}_r ]$, in which the equation $\text{E}[\alpha \alpha^*]=P_R$ is used.

From (\ref{y_dist}), it can be seen that $\boldsymbol{y}^{(s,K^2)}$ follows a zero mean, circularly symmetric complex Gaussian (CSCG) distribution. Using the conditional co-variance from (\ref{sigma}) in (\ref{y_dist}), the corresponding conditional probability density function (PDF) is defined as \cite{gallager2008circularly}
\begin{align} \label{f_pdf}
    f(\boldsymbol{y}^{(s,K^2)}&|\hat{k}_t,\hat{k}_r) =\\ & \frac{1}{\pi^{K^2}\text{det}(\boldsymbol{\Sigma}_{(\hat{k}_t,\hat{k}_r)})} \text{exp}(-(\boldsymbol{y}^{(s,K^2)})^H \boldsymbol{\Sigma}^{-1}_{(\hat{k}_t,\hat{k}_r)} {\boldsymbol{y}^{(s,K^2)}}). \nonumber
\end{align}
Now let us find the conditional probability of $\hat{k}_t$ and $\hat{k}_r$ given the receive measurement vector $\boldsymbol{y}^{(s,K^2)}$. Define ${\mathcal{K}}= \{ \{ 1,1 \},\{ 1,2 \}, \cdots , \{ K,K \} \}$ as the set of all $K^2$ possible sub-range combinations such that $\{ \hat{k}_t,\hat{k}_r \} \in {\mathcal{K}}$. For simplicity, we assume that each sub-range combination has an equal probability of having an AOD/AOA pair. Following the principle of maximum likelihood detection, we can express the probability of $\hat{k}_t$ and $\hat{k}_r$ given $\boldsymbol{y}^{(s,K^2)}$ as
\begin{align} \label{sub_range_prob}
   p(\hat{k}_t&,\hat{k}_r|\boldsymbol{y}^{(s,K^2)})  =  \frac{ f(\boldsymbol{y}^{(s,K^2)}|\hat{k}_t,\hat{k}_r)  }{ \sum\limits_{\{\hat{k}_t',\hat{k}_r'\} \in {\mathcal{K}} }  f(\boldsymbol{y}^{(s,K^2)}|\hat{k}_t',\hat{k}_r')   }. 
\end{align}
\noindent
The most likely sub-range combination can then be obtained as
\begin{align}
\label{sub_range_est}
\{ \hat{k}_t&,\hat{k}_r \}=\underset{{\{\hat{k}_t,\hat{k}_r\} \in {\mathcal{K}} }}{\operatorname{argmax }} [  \; p \big(\hat{k}_t,\hat{k}_r|\boldsymbol{y}^{(s,K^2)} \big) \; ].
\end{align}
These estimates can then be used to reduce the ranges of possible AOA and AOD to, respectively, the $\hat{k}_t$th transmit and $\hat{k}_r$th receive angular sub-ranges for estimation in the next stage. Upon completing the final stage of estimation, the channel coefficient $\alpha$ can be estimated from the previously obtained measurements as
\begin{align}
\label{alpha_est}
\hat{\alpha} =  \frac{  \sqrt{P} x (\boldsymbol{w}_{\hat{k}_r}^{(S)})^H \boldsymbol{H} \boldsymbol{f}_{\hat{k}_t}^{(S)} }{\sqrt{P}N C^2_{(S)} ||x||_2}.
\end{align}

It is important to emphasize the significance of correct estimation in the early stages of estimation. It should be noted that the directivity gain will become larger in each subsequent stage. This is because the reduced angular sub-ranges to be estimated require narrower beam patters for the later stages. However, if any estimate in the prior stage was wrong and the angular subranges were determined incorrectly, the error cannot be corrected in a later stage, resulting in a final incorrect AOD/AOA estimate. As earlier stages have lower directivity gain, this is also where the errors are most likely to occur. In the following subsection, we will overcome this problem by developing an adaptive algorithm to dynamically allocate more measurements to stages that expect a higher probability of error.

\subsection{Rate Adaptive Channel Estimation Algorithm}

An important function of the ML estimator developed in the previous subsection is the insight into the probability defined in (\ref{sub_range_prob}). Generally, multi-stage channel estimation schemes such as \cite{rheath} use a fixed number of channel measurements in each stage. Particularly, the estimator at the receiver is forced to make a decision after $K^2$ measurements, without taking into account the probability $p(\hat{k}_t,\hat{k}_r|\boldsymbol{y}^{(s,K^2)})$. Leveaging the tools developed in the previous subsection, we now propose to estimate the channel state information by transmitting a potentially infinite number of pilot signals and beam patterns. As existing multi-stage mmWave channel estimation schemes already rely on feedback, we use this feedback to assist in the generation of each new beam pattern, by re-measuring on the most likely sub-range estimation pair. By adopting such an approach, the number of measurements required for channel estimation is able to dynamically adapt to the channel conditions.

To describe our proposed algorithm, we first introduce a target maximum probability of estimation error (PEE), denoted by $\Gamma$. The basic principle of the RACE algorithm is that, after the $m=K^2$ initial measurements are completed in any given stage, if the most likely sub-range pair does not satisfy $p(\hat{k}_t,\hat{k}_r|\boldsymbol{y}^{(s,m)})>(1-\Gamma)$, then additional measurements will be performed. To this end, the receiver will still feedback the current most likely transmit sub-range, $\hat{k}_t^{(s)}$ as it does in \cite{rheath}, however it will also feedback the result of whether more measurements are required or not. This leads to, at most, one additional bit of information per feedback compared to \cite{rheath}. More specifically, we require $\lceil\text{log}_2(K)+1\rceil$ feedback bits, which will be shown to be negligible at high SNR as the average number of additional measurements converges to zero.

In the case where the specified probability threshold was not met after the $m$th measurement, instead of further dividing the sub-ranges corresponding to $\hat{k}_t^{(s)}$ and $\hat{k}_r^{(s)}$, an additional measurement is taken on this sub-range combination, using beamforming vectors $\boldsymbol{f}_{\hat{k}_t}^{(s)}$ and $\boldsymbol{w}_{\hat{k}_r}^{(s)}$. The updated received measurement vector after the $m+1$ time slots can then be expressed by appending the new measurement to the previous measurements as
\begin{align}
\label{y_m_1}
\boldsymbol{y}^{(s,m+1)} =& \sqrt{P} x \left[\begin{array}{ccc} \boldsymbol{h}_v^{(s,m)} \\ (\boldsymbol{w}_{\hat{k}_r}^{(s)})^H  \boldsymbol{H} \boldsymbol{f}_{\hat{k}_t}^{(s)} \end{array}\right] +\left[\begin{array}{ccc} \boldsymbol{n}^{(s,m)}  \\ (\boldsymbol{w}_{\hat{k}_r}^{(s)})^H \boldsymbol{q}_{m+1}   \end{array}\right].
\end{align}
The estimated sub-ranges and corresponding probability can then be updated by substituting $\boldsymbol{y}^{(s,m+1)}$ into (\ref{sub_range_prob}-\ref{sub_range_est}). 

\begin{figure}[!t]
\centering
\includegraphics[width=3.0in,trim={0.1cm 4.50cm 2.9cm .4cm},clip]{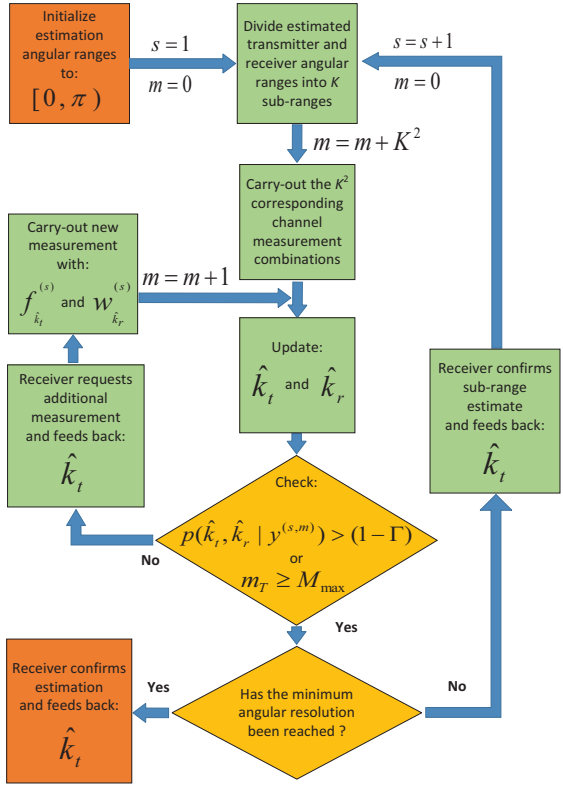}
\caption{Flow Diagram of the Proposed RACE Algorithm.}
\label{race_flow_diagram}
\end{figure}  

We then repeat this process until either the threshold condition has been met (i.e., $p(\hat{k}_t,\hat{k}_r|\boldsymbol{y}^{(s,m)})>(1-\Gamma)$) or a maximum number of measurements, denoted by $M_{\max}$, has been reached (i.e., $m_T \geq M_{\max}$, where $m_T$ is the total number of measurements carried out in all stages). Under fading channel conditions, there always exists a non-zero probability of an ‘outage’ occurring when the path coefficient is close to zero. By imposing an upper limit to the number of measurements, we reduce the time and energy expended in this case. A complete flow diagram of the RACE algorithm is shown in Fig. \ref{race_flow_diagram}. As the RACE algorithm is able to compute the probability of successful estimation during the estimation process, it can minimize the number of measurements required for estimation and therefore reduce the required energy.

\subsection{A Rate-Switching Benchmark Design}

We now aim to fairly compare our proposed RACE algorithm to another scheme that can also change its estimation rate. To do so, we propose a generalization of the algorithm in \cite{rheath}, such that each stage $s$ is able to use an arbitrary number of beam patterns denoted by $K_s$. We then define the vector describing the number of beam patterns used in all stages as $\boldsymbol{K}=[K_1,...,K_S]$. The angular resolution reached after the $i$th stage of estimation is then given by $\frac{\pi}{\prod_{s=1}^i K_s}$. In order to reach the minimum angular resolution after $S$ stages, an important property of $\boldsymbol{K}$ is then that $\frac{\pi}{N} = \frac{\pi}{\prod_{s=1}^S K_s} $, i.e., the sub-range division in all stages leads to the minimum angular resolution. 
We can then describe the total number of estimation measurements for any $\boldsymbol{K}$ as $M_T= \sum_{s=1}^S K_s^2 $. 

Based on this generalization, we then propose that if the PEE exceeds a predefined threshold, the number of beam patterns used in the first stage is increased. For example, consider the channel estimation of an $N_t=N_r=8$ antenna system using $\boldsymbol{K}=[2,2,2],$ which requires $S=3$ stages with $M_T = 12$ total measurements. If the PEE is predicted to exceed a predetermined threshold, the proposed benchmark approach switches to instead use $\boldsymbol{K}=[4,2]$, leading to an improved PEE owing to the first stage directivity gain. The cost for this improvement is a much greater number of measurements in the first stage and a higher total number of measurements with $M_T = 20$. In the simulation results presented in next section, we refer to this approach as rate switching.  

\begin{figure*}[!t]
\centering
\subfigure[]{\includegraphics[width=3.5in,trim={0.75cm 6.8cm 1.5cm 8.7cm},clip]{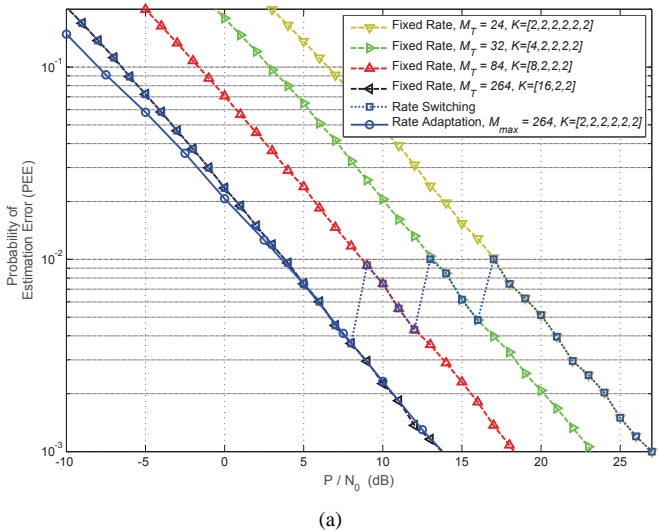}}
\subfigure[]{\includegraphics[width=3.5in,trim={0.75cm 6.9cm 1.5cm 8.7cm},clip]{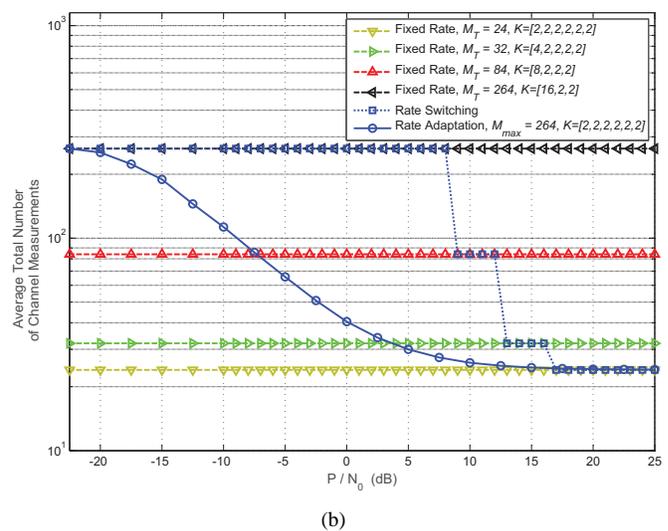}}
\caption{Numerical performance results of the proposed algorithms compared to the algorithm presented in \cite{rheath} for (a) probability of estimation error (PEE) and (b) average number of measurements required for channel estimation.}
\label{numerical_res}
\end{figure*}  

\section{Numerical Results}
We now provide some numerical results to illustrate the performance of our proposed RACE algorithm. We consider a mmWave system with $N = 64$ antennas at both the transmitter and receiver. We use a single path channel with fading coefficient, $\alpha$, assumed to follow a complex Gaussian distribution with zero mean and variance $P_R=1$. We assume that the corresponding AOD, $\phi^t \in \mathcal{U}_N $, and AOA, $\phi^t \in \mathcal{U}_N $,  follows a random uniform distribution. We compare our results with the multi-stage algorithm in \cite{rheath} using different values of $K$ in each stage. More specifically, by employing larger $K$ in earlier stages, greater beamforming gains can be achieved, reducing the probability of channel estimation error at early stages. However this leads to a greater total number of measurements overall. In our proposed algorithm we use $K = 2$ in all stages, resulting in a minimum of $K ^2= 4$ measurements in each stage. We use a target PEE of $\Gamma=10^{-2}$ and the number of maximum measurements $M_{\max}$ is set to be the maximum number of measurements used by the comparison schemes of \cite{rheath}, i.e., $M_{\max}=264$. All schemes use a fixed transmit power, $P$, in all measurements.

Fig. \ref{numerical_res}(a) shows the probability of an incorrectly estimated AOD/AOA pair after the $S$ stages of estimation have been carried out and Fig. \ref{numerical_res}(b) shows the average total number of measurements required for the same estimation. We can see that the RACE algorithm is able to achieve similar performance to the $M_T=264$ fixed rate scheme (i.e., the scheme requiring the most measurements) but requires significantly less measurements across a large range of SNR. Focussing on higher SNR, we can see that from $P/N_0>12.5$ dB the RACE algorithm converges on the same average number of measurements as the $M_T=24$ fixed rate scheme from \cite{rheath} (i.e., the scheme requiring the least measurements) but achieves significantly better performance. We also show a rate switching approach which is able to switch between each of the fixed rate schemes in attempt to also keep the error probability below $\Gamma=10^{-2}$. In general it can be seen that, the proposed RACE algorithm can achieve significant performance gains with greatly reduced number of channel measurements, compared to existing fixed rate or rate switching schemes. 


\section{Conclusion}

In this paper we proposed a novel rate adaptive channel estimation for mmWave MIMO communication systems. In the proposed algorithm, additional measurements are carried out when a high probability of estimation error is expected. This was achieved by elaborately designing a maximum likelihood estimator, which can optimally estimate the channel information and the associated probability of estimation error. We have shown that the proposed approach yields a similar probability of channel estimation error as the best comparison algorithm from existing methods but requires a significantly lower number of measurements at high SNR.


\bibliography{IEEEabrv,Globecom_2015}

\end{document}